\documentclass[twocolumn]{aastex631}
\usepackage{CJK}

\shorttitle{Sub-Alfv\'enic Solar Wind Fluctuations}
\shortauthors{Ruffolo et al.}
\graphicspath{{./}{figures/}}

\begin{document}
\begin{CJK*}{UTF8}{gbsn}

\title{
Observed Fluctuation Enhancement 
and  Departure from WKB Theory
in Sub-Alfv\'enic Solar Wind}

\correspondingauthor{David Ruffolo}
\email{david.ruf@mahidol.ac.th}

\author[0000-0003-3414-9666]{David Ruffolo}
\affiliation{Department of Physics, Faculty of Science, Mahidol University, Bangkok 10400, Thailand}

\author[0000-0001-9597-1448]{Panisara Thepthong}
\affiliation{Department of Physics, Faculty of Science, Kasetsart University, Bangkok 10900, Thailand}

\author[0000-0002-6609-1422]{Peera Pongkitiwanichakul}
\affiliation{Department of Physics, Faculty of Science, Kasetsart University, Bangkok 10900, Thailand}

\author[0000-0003-3891-5495]{Sohom Roy}
\affiliation{Department of Physics and Astronomy, University of Delaware, Newark, DE 19716, USA}

\author[0000-0003-4168-590X]{Francesco Pecora}
\affiliation{Department of Physics and Astronomy, University of Delaware, Newark, DE 19716, USA}

\author[0000-0002-6962-0959]{Riddhi Bandyopadhyay}
\affiliation{Department of Astrophysical Sciences, Princeton University, Princeton, NJ 08544, USA}

\author[0000-0002-7174-6948]{Rohit Chhiber}
\affiliation{Department of Physics and Astronomy, University of Delaware, Newark, DE 19716, USA}
\affiliation{Heliophysics Science Division, NASA Goddard Space Flight Center, Greenbelt, MD 20771, USA}

\author[0000-0002-0209-152X]{Arcadi V. Usmanov}
\affiliation{Department of Physics and Astronomy, University of Delaware, Newark, DE 19716, USA}
\affiliation{Heliophysics Science Division, NASA Goddard Space Flight Center, Greenbelt, MD 20771, USA}

\author[0000-0002-7728-0085]{Michael Stevens}
\affiliation{Center for Astrophysics, Harvard \& Smithsonian, Cambridge, MA 02138, USA}

\author[0000-0002-6145-436X]{Samuel Badman}
\affiliation{Center for Astrophysics, Harvard \& Smithsonian, Cambridge, MA 02138, USA}

\author[0000-0002-4559-2199]{Orlando Romeo}
\affiliation{Department of Earth \& Planetary Science and Space Sciences Laboratory, University of California at Berkeley, Berkeley, CA 94720, USA}

\author[0009-0008-8723-610X]{Jiaming Wang}
\affiliation{Department of Physics and Astronomy, University of Delaware, Newark, DE 19716, USA}

\author[0009-0008-8723-610X]{Joshua Goodwill}
\affiliation{Department of Physics and Astronomy, University of Delaware, Newark, DE 19716, USA}
\author[0000-0002-5317-988X]{Melvyn L. Goldstein}
\affiliation{Space Science Institute, Boulder, CO 80301, USA}

\author[0000-0001-7224-6024]{William H. Matthaeus}
\affiliation{Department of Physics and Astronomy, University of Delaware, Newark, DE 19716, USA}
\affiliation{Bartol Research Institute, University of Delaware, Newark, DE 19716, USA}


\begin{abstract}
 Using Parker Solar Probe data from orbits 8 through 17,
 we examine fluctuation amplitudes
throughout the 
critical region where the solar wind flow speed approaches and then exceeds the Alfv\'en wave speed, taking account of various 
exigencies of the plasma data.  
In contrast to WKB theory for non-interacting Alfv\'en waves streaming away from the Sun, 
the magnetic and kinetic fluctuation 
energies per unit volume 
are not monotonically decreasing.
Instead, there is clear violation of conservation of standard WKB wave action, which is consistent with previous 
indications of strong {\it in situ} fluctuation energy input in the solar wind near the Alfv\'en critical region. 
This points to strong violations of WKB theory due to nonlinearity (turbulence)  and major energy
input near the critical region, which we interpret as likely due to driving by large-scale coronal shear flows.
\end{abstract}

{\hfill Submitted for publication in the {\it Astrophysical Journal Letters}, Aug.\ 3, 2024}

\section{Introduction} 
\label{sec:intro}
\end{CJK*}

An important
historical view of the fluctuations observed in the solar wind, dating to the early days of space exploration, is that 
these are predominantly ``fossil'' Alfv\'en waves that propagate outward from the Sun with little interaction
\citep{Coleman67wave,BelcherDavis71}. 
This places specific constraints on the 
properties of observed 
magnetic and velocity fluctuations \citep{Hollweg74,Barnes79a}
and greatly simplifies many 
models of the interplanetary transport of fluctuations 
\citep{Hollweg90,VermaRoberts93,SquireEA20,FiskKasper20}.
Frequently 
the assumption of
unidirectional propagation
goes along with an assumption of small amplitude transverse 
Alfv\'en waves, often considered ``slab-like'' in that they vary along one coordinate direction.
In the case that the wavelength is small compared to the scale of variation of 
a weakly 
inhomogeneous background, 
the transport of small-amplitude (non-interacting) Alfv\'en waves
is governed by WKB theory
\citep{Weinberg62,Parker65-ssr}.
The
present paper delves
into specific details regarding the applicability of WKB theory in the solar wind. 

The transport of one-dimensional waves having arbitrary wave lengths
was considered by
\cite{HeinemannOlbert80}.
Although this model is much simpler than the three dimensional (3D)
model for turbulence transport
\citep{ZhouMatt89-nonwkb,MattEA94-ec},
it suffices to demonstrate that 
violations of WKB ordering
produce coupling between 
inward- and outward-propagating fluctuations even when nonlinear 
effects are neglected.
A more 
complete 3D transport model \citep{VermaRoberts93} 
permits 
varying cross-helicity and 
nonequipartition of velocity and magnetic fluctuation energy,
but still
neglects
energy input and 
turbulent dissipation;
this model found that 
the radial 
variation of fluctuation
energy departed only marginally from WKB solutions. 
An analysis of a broad range
of solutions 
to
{\it linear} transport equations for solar wind fluctuations
was presented by \citet{OughtonMatt95}. 
Later \citet{ZankEA96}
showed that a full set
of turbulence transport equations including 
shear driving and von Karman MHD dissipation \citep{WanEA12-jfm}
also closely follows 
WKB radial energy profiles, whereas the same equations without shear driving do not. 
This indicates that observation of WKB-like radial energy profiles 
in the (super-Alfv\'enic) solar wind 
is not sufficient to conclude that WKB theory is valid.
In fact, beyond this 
single issue, for super Alfv\'enic wind far outside the critical region, e.g., at 1 au, 
there are numerous 
observational findings that demonstrate convincingly that the assumptions and (other) consequences of WKB theory are inconsistent with 
observations \citep{MatthaeusVelli11}.

The validity of WKB theory for the sub-Alfv\'enic corona is much less well 
determined in the existing 
literature. 
This paper deals with that 
issue.

On the theoretical side,
the presence of a 
large Alfv\'en speed 
would seem to favor wave propagation effects and therefore the physics 
entailed by WKB theory. 
Indeed WKB and its close relatives \citep{HeinemannOlbert80}
are often involved in descriptions of 
coronal phenomena \citep[see, e.g.,][]{RaouafiEA23-parker}.
Fortunately the Parker Solar Probe mission,
with its regular forays into the subAlfv\'enic corona \citep{ChhiberEA24}, 
provides for the first time
an opportunity
for direct observational testing 
of the relevance of WKB theory below the critical region.

\section{Data and Analysis Procedure} \label{sec:data}
We employ 
Parker Solar Probe (PSP) data, 
mainly derived from publicly accessible 
data archives.
We use magnetic field ($\bf{B}$) data from the flux-gate magnetometer of the PSP/FIELDS instrument suite \citep{Bale_2016}\footnote{ http://research.ssl.berkeley.edu/data/psp/data/sci/fields/l2/}, downsampled to a frequency of 4 Hz.  
Solar wind proton velocity ($\bf{V}$) 
data obtained from moments of the particle velocity data, are from the Solar Probe ANalyzer for Ions (SPAN-i) instrument in the PSP/SWEAP instrument suite \citep{Kasper_2016}\footnote{http://sweap.cfa.harvard.edu/pub/data/sci/sweap/}.  
We use SWEAP data at the native cadence, which is most commonly 0.87 s, or 1/4 of that near perihelion passages.

In our analysis 
we address two major concerns with the quality of the
SWEAP/SPAN datasets: 
1) The SPAN-i instrument has a limited
field of view \citep{Kasper_2016,Livi2022} meaning it often has incomplete sampling of proton velocity distribution functions (VDFs). 
The field of view constraints are most severe in the tangential direction \citep[or close to $\varphi$ in instrument coordinates; see Figure 2 of][]{Livi2022}.
Nevertheless, if the peak of the VDF falls into the instrument field of view,
then the velocity moment vector is typically quite accurate. 
We filter the velocity moment data via the `EFLUX\_VS\_PHI' CDF variable in the SPAN-i L3 data. 
If the peak energy flux is located lower than $160^\circ$ in instrument azimuthal coordinates, then the velocity moment is accepted as a ``good measurement''. 
This azimuthal cutoff is conservative in that it also rejects measurements where the shadow of the PSP spacecraft heat-shield may distort the peak location.

2) 
Due to the above concerns,
and after considerable experimentation
with different approaches,
we adopt  a hybrid procedure
to secure required plasma (proton) number density
measurements $N_p$.
First, when it is available
we associate $N_p$ 
with the electron number density ($n_e$)
obtained from quasi-thermal 
noise (QTN) electric field data 
from the FIELDS suite \citep{Moncuquet_2020}. We adopt a simplified heuristic approach by \citet{romeo2023} to determine the QTN electron number density for each orbit. This method calculates the number density within $5-10\%$ of the density estimates derived by \citet{Moncuquet_2020}.
We then perform 
a final stage of filtering to achieve a continuous signal.
An exception is that for PSP solar encounter E15, we use the QTN data directly from the Moncuquet et al.\ procedure\footnote{https://research.ssl.berkeley.edu/data/psp/data/sci/fields/l3/rfs\_lfr\_qtn/}.
For all of the PSP encounters considered here, E8 through E17, when
 QTN data are not available,
we use $n_p$
data from SPAN, provided that 
the criterion
regarding resolution
of the particle distribution
is satisfied. 
The measured 
SPAN $n_p$ is then multiplied by an empirical factor of 0.86
(derived by a comparison with $n_e$ data at the times when SPAN is in the field of view) to arrive at a plasma density $N_p = 0.86 n_p$. 
When available from either 
of these sources,
the useful plasma density 
data $N_p$ is downsampled (averaged) to 1-minute cadence. This provides a times series with resolution several times smaller than the typical correlation times during these encounters, 
thus enabling diagnostics that measure properties of the local turbulent fluctuations. 



For each minute of the 
data, selected as described above, we derive an 
Alfv\'en speed, $V_A$, of the proton-dominated plasma 
from 
\begin{equation}
V_A = \frac{\langle|\bf{B}|\rangle}{\sqrt{\mu_0m_p\langle N_p\rangle}},
\label{eq:V_A}
\end{equation}
where $\langle|\bf{B}|\rangle$ is the mean magnitude of the magnetic field during that minute, $\langle N_p\rangle$ is the mean 
plasma (proton) number density, as described above, during that minute, and $m_p$ is the proton mass.
The Alfv\'en Mach number $M_A$ is then calculated as 
\begin{equation}
    M_A = \frac{\langle V_R\rangle}
              {V_A}
              \label{eq:AlfvenMach},
\end{equation}
the ratio of the mean proton radial velocity  to the local Alfv\'en speed for that minute.

Also required for this analysis 
are derived measures of turbulent fluctuations, especially 
$\delta B$ for the rms magnetic fluctuation.
We use this case as an example. 
For each minute of the data 
${\bf B}$, 
we find the mean magnetic field components $\langle B_i\rangle$.
Then we determine $(\delta B_i)^2
= \langle B_i^2\rangle - \langle B_i  \rangle^2 $ 
as the mean-squared fluctuation of each field component relative to its mean value, i.e., the variance of that component during the minute. 
Finally, 
$\delta B = \sqrt{\sum_{i=1}^3(\delta B_i)^2}$.
An analogous procedure was used to determine the velocity component fluctuations $\delta V_i$ for each minute of data.  

Incorporating the 
proton
density dataset prepared according to the 
above procedure, 
and in consultation with 
various members of the 
SWEAP instrument team, 
we prepared an updated
time series of 
Alfv\'en speed and Alfv\'en 
Mach number (Eq. \ref{eq:AlfvenMach}).
This dataset
differs from typical simpler
analyses, including that used in our own recent work
\citep{ChhiberEA24}.
Fortunately the differences are, on balance, fairly subtle, 
consisting mainly of nearly isolated short periods wherein use of the more primitive SPAN dataset produces ``spikes'' of low Alfv\'en Mach number
due to density dropouts at heliodistances greater than \(\sim 30 ~R_\odot\).
Most of these
disappear when the
more completely analyzed density data is employed.
Our compilation of PSP density data will be made public. 

Using the refined Mach number data allows 
examination of 
scalings of magnetic fluctuations with respect to $M_A$ and with respect 
to radius $r/R_\odot$ 
to be carried out for 
numerous PSP orbits. 
Here we show data from encounters E8 through E17. 
We also examine velocity fluctuation energy per unit volume by computing a closely related surrogate $\rho \delta V^2 \equiv \rho \delta V_R^2 + 2 \rho \delta V_N^2$. 
Here $\delta V^2$ is computed from the radial velocity component variance $\delta V_R^2$ and {\it twice} the  
normal component variance $\delta V_N^2$, i.e.,
using $\delta V_N^2$ as a proxy for $\delta V_T^2$. 
This avoids use of the tangential velocity component, whose distribution is cut off in the +T direction by SPAN field of view effects \citep{BadmanEA2023AGU}.
The use of the surrogate amounts to asserting that the velocity field fluctuations are 
axisymmetric about the radial direction \citep{OughtonEA15}. 
The assumption of axisymmetric fluctuations is supported by observations of magnetic fluctuations, e.g., in the present data set the average of $\delta B_N^2/\delta B_T^2$ is within 1\% of unity, and previous observations from Mariner 4 \citep{Belcher_Davis_1971} and from PSP 
\citep{Ruffolo2020,Fargette22,Chhiber22} found that tangential-like and normal-like perpendicular increments of $\mathbf{B}$ have similar variances, especially at lags of 60 s or less.

\begin{figure*}[t]

\includegraphics[width=\linewidth]{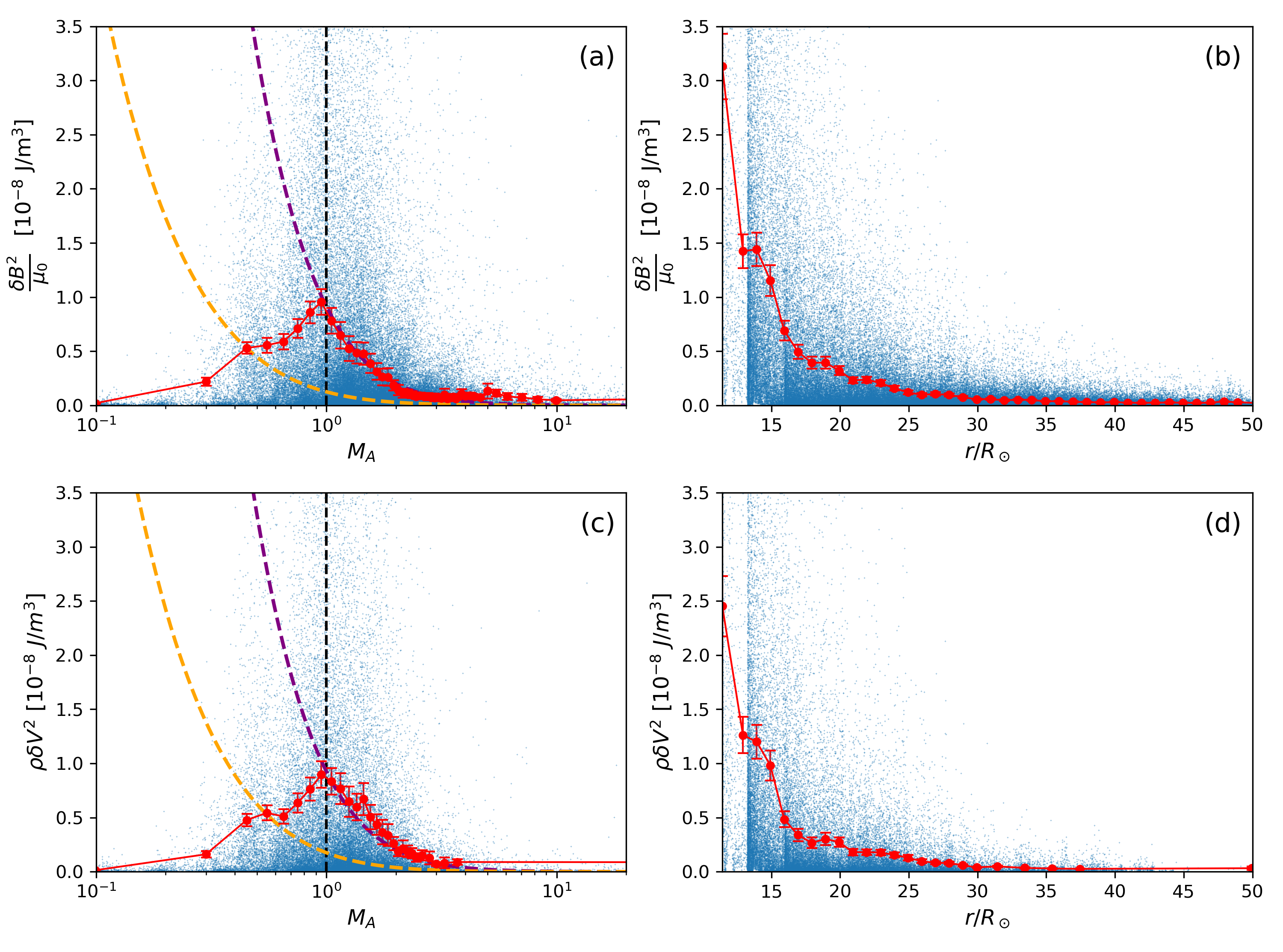}
\caption{
Solar wind fluctuation energy per unit volume 
vs.\ Alfv\'en Mach number $M_A$
and radial distance $r$ for PSP data during orbits 8 to 17. 
Each blue point shows a 1-minute-averaged value. Red circles indicate the average value in each bin of $M_A$ having at least 100 data points, and the extent of vertical red lines indicates the standard deviation of data in that bin. Magnetic fluctuation energy per volume 
(a) $\delta B^2/\mu_0$ vs.\ $M_A$; and (b) vs.\ $r/R_\sun$.
Velocity fluctuation energy per volume
(c) $\rho \delta V^2$ vs.\ $M_A$; and (d) vs.\ $r/R_\sun$.
$M_A$ is computed using the density data set described in the text. 
Velocity fluctuation is computed only when SPAN-i 
data satisfy quality conditions described in the text. 
The evolutionary trend predicted from WKB theory for ``fossil'' Alfv\'enic fluctuations that propagate outward without interaction, ${\cal E}\propto [M_A(M_A+1)^2]^{-1}$ (dashed curves in panels (a)  and (c)), fails to explain the PSP data for sub-Alfv\'enic solar wind, i.e., for $M_A<1$.
The discrepancy implies that most solar wind fluctuation energy does not originate near the solar surface but rather is strongly enhanced {\it in situ} at $0.5\lesssim M_A\lesssim1$.
}
   \label{fig:db-MA-02}
\end{figure*}

\section{Results: Fluctuation Scaling 
versus Alfv\'en Mach number} \label{sec:results}

Employing the data processing procedures outlined above, 
we have in hand data from ten PSP orbits that are suitable for 
quantifying the level of adherence to 
WKB theory expectations. 

Central to this question are the 
variations of the solar wind fluctuation energy densities
per unit volume; for convenience, we suppress the factor of two in the definition of energy density, using $\delta B^2/\mu_0$ or $\rho\delta V^2$. The evolution of energy density is shown
in Figure \ref{fig:db-MA-02}.
The top row is for the magnetic fluctuation energy and the bottom row is for flow velocity fluctuation energy.
Alfv\'en Mach dependence is shown in the left column, 
and radial distance dependence on the right. 
The points are widely scattered but bin averages, shown by 
connected symbols (red), provide well-defined tendencies.

In the evolution of the solar wind, a basic expectation is that $M_A$ increases with increasing distance $r$ from the Sun, as the wind changes from sub-Alfv\'enic to super-Alfv\'enic.  
However, in Figure \ref{fig:db-MA-02}, there is a strong visual difference between the dependence of various quantities versus $M_A$ or $r/R_\sun$.
Partly this is because different streams of the solar wind evolve differently in $M_A$ as a function of $r$, and according to global MHD simulations that include turbulence, $M_A$ can be non-monotonic even for a single stream \citep{ChhiberEA22}.
Note that the implementation of the WKB approximation \citep{Parker65-ssr,Barnes75} assumes a steady state, a radial mean field, and $B_0\propto r^{-2}$, but does not specify the radial dependence of the density and therefore retains the flexibility that different solar streams can evolve similarly in $M_A$ but differently versus $r$.
Therefore, the WKB predictions for fluctuation energy density ${\cal E}$ can be expressed in terms of $M_A$, in particular as ${\cal E}\propto[M_A(M_A+1)^2]^{-1}$ \citep{Jacques78}, and not in terms of $r$ unless additional assumptions are imposed.
In this sense, the dependence of solar wind fluctuation parameters on $M_A$ is expected to be fundamental, and can be obscured when expressed in terms of $r$ because of the different evolution of different solar wind streams.
In Figure \ref{fig:db-MA-02}(a) and (c), the Alfv\'en Mach number dependence expected from standard WKB theory
is suggested in dashed trend lines versus $M_A$, 
chosen to intersect the observed average energy per volume at values that facilitate comparison and discussion.

A familiar and not unexpected  behavior 
of $\delta B^2$ versus $r/R_\sun$ 
is seen in 
Figure \ref{fig:db-MA-02}(b); 
it is essentially monotonically
decreasing. 
However the 
behavior of $\delta B^2$ versus $M_A$ in
Figure \ref{fig:db-MA-02}(a)
is less familiar in existing theoretical work on wave propagation and is
not anticipated in WKB theory. 
In fact the WKB theory of fossil, non-interacting, outgoing Alfv\'enic fluctuations predicts a monotonic decrease of $\delta B^2$ as a function
of $M_A$, with $\delta B^2\propto [M_A(M_A+1)^2]^{-1}$ as represented by the reference traces (dashed curves) in 
the Figure, which is not obtained for the PSP data as analyzed here. 

Two rather distinct types of behavior are seen, separated 
by $M_A = 1$. Under sub-Alfv\'enic 
(coronal) conditions, with 
$M_A < 1$, one sees in the bin averages of $\delta B^2$ 
a regular increase when moving towards the Alfv\'en critical point $M_A = 1$. 
In the same region,  the 
WKB reference traces are steeply declining. 
Since an inner 
boundary value for the WKB
traces is arbitrary here, we chose 
the upper trace in panel (a) 
so that the reference curve coincides with the value of
$\delta B^2/\mu_0$ at $M_A=1$.
It is then immediately apparent that the bin averaged 
$\delta B^2$ at superAlfv\'enic values of $M_A$ actually are quite well fit by the 
WKB profile. 
As striking as this agreement is, it is also not at all a surprise, since WKB-like behavior of $\delta B^2$
has been noted in data at 1 au in turbulence transport calculations when 
both shear driving and 
dissipation are included 
\citep{ZankEA96}, and even in simpler models 
\citep{VermaRoberts93}.
However the extension of 
the curve that agrees so well with WKB at $M_A > 1$ is dramatically different from the averaged data values in the sub-Alfv\'enic region. 
Another WKB profile, the lower dashed curve in panel (a), passes through the 
bin averaged $\delta B^2$ data
at around $M_A = 0.5$, but clearly fails to represent any aspect of the magnetic energy per unit volume in the sub-Alfv\'enic region.

Turning attention to the 
second row of panels 
in Figure \ref{fig:db-MA-02}, the quantity of interest changes to 
the kinetic energy per unit volume in the velocity fluctuations. 
The right panel (d) 
exhibits an essentially
continuously decreasing 
value of $\rho \delta V^2$ 
as a function of distance $r/R_\sun$, 
qualitatively 
quite similar to 
the magnetic energy density in panel (b).
The behavior of 
kinetic energy per volume 
$\rho \delta V^2$ vs.\ 
Alfv\'en Mach number
also is very similar to 
that of $\delta B^2$ in panel (a): 
once again there are 
gross departures from WKB theory in the sub-Alfv\'enic coronal region, 
while the 
functional behavior in the super-Alfv\'enic region overlays nearly as well for $\rho \delta V^2$ vs.\ $M_A$ as it did in panel (a) for $\delta B^2$ vs.\ $M_A$. 


\begin{figure}[h]
\includegraphics[width=\columnwidth]{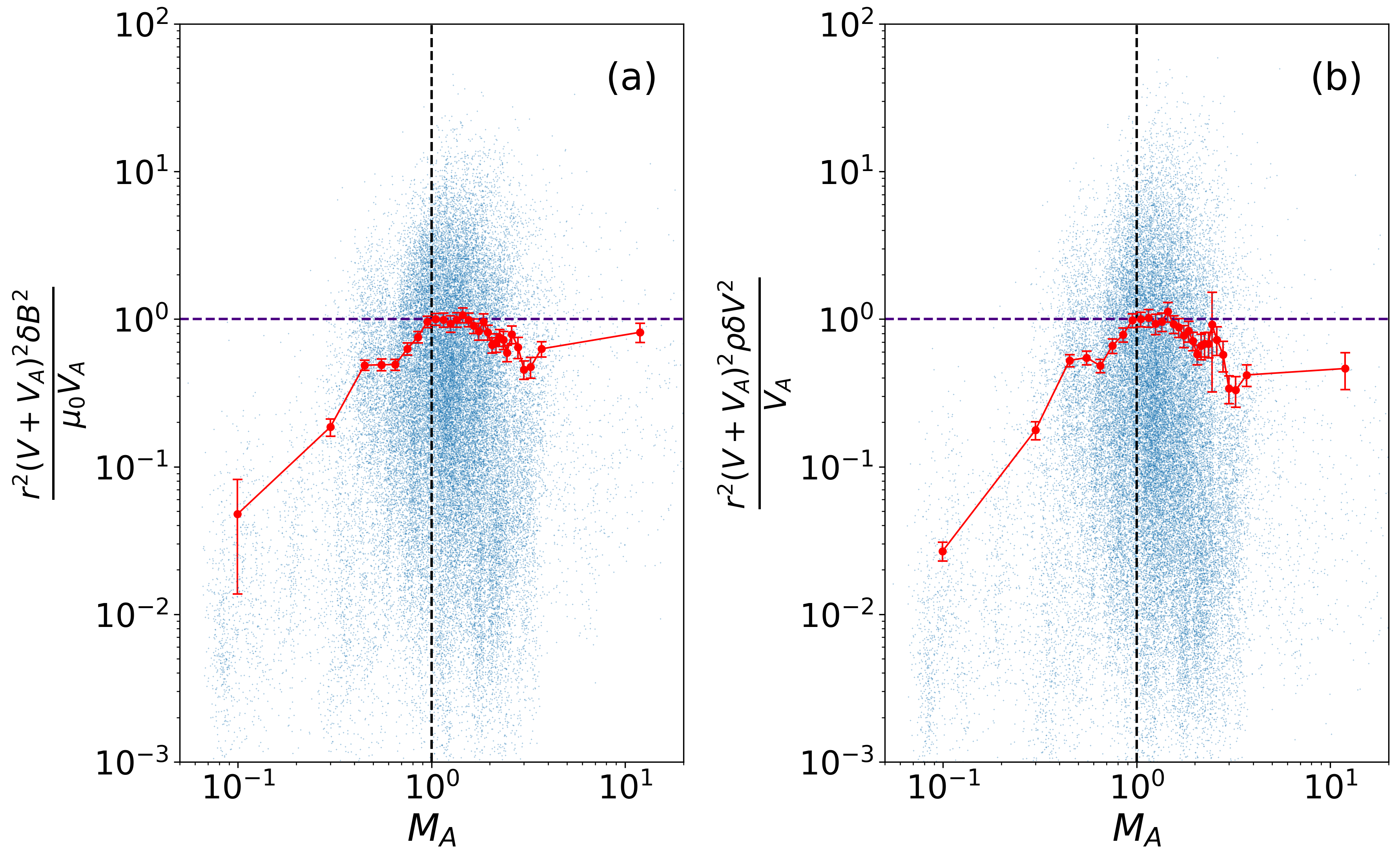}
\caption{
Quantities related to Alfv\'en wave action 
vs Alfv\'en Mach number $M_A$, 
expected to be 
conserved in WKB theory.
(a) $r^2 (V + V_A)^2 \delta B^2 / (\mu_0 V_A)$
vs $M_A$, and
(b) $r^2 (V + V_A)^2 \rho \delta V^2 / V_A$ vs $M_A$, evaluated here 
from PSP data, are 
not constant as function of $M_A$, except possibly for $1<M_A<2$.
Solid (red) line is bin-averaged data. Vertical axes are normalized to the value at $M_A=1$ (horizontal dark blue line).
As a guide, 
vertical black dashed line at $M_A = 1$ indicates 
the Alfv\'en transition zone.
In the sub-Alfv\'enic region ($M_A<1$) these quantities are increasing,
violating conservation of wave action and suggesting {\it in situ} fluctuation energy enhancement.}
   \label{fig:wave_energy}
\end{figure}

The evidence so far presented from PSP 
depicts significant departures from WKB 
profiles.
Another perspective on this is obtained by examining the 
behavior of the 
{\it wave action}
in a familiar standard form.
In particular, assuming a radial mean field, non-interacting Alfv\'en waves in a weakly inhomogeneous medium, the conservation of mass flux and magnetic flux, and the absence of damping,
one finds that \citep{Jacques77}
\begin{equation}
\frac{r^2(V+V_A)^2 {\cal E}}{V_A} = const.
\label{eq:action}
\end{equation}
where  ${\cal E} = \rho \delta V^2 = \delta B^2/\mu_0$ according to the equipartition of velocity and magnetic fluctuations
for unidirectionally propagating Alfv\'en waves.
This relation 
is usually called conservation of wave action, and
is a central property of standard WKB as applied to the solar wind  
\citep{Parker65-ssr,Barnes75}.
The same collection of data used above can be 
employed to evaluate the degree 
to which the wave action so defined
is conserved in the PSP observations. 
This analysis is shown in Figure \ref{fig:wave_energy}, 
showing scatter plots of the left hand side of 
Eq. (\ref{eq:action})
for ${\cal E} = \delta B^2/\mu_0$ (left panel) 
or $\rho \delta V^2$ (right panel),
in each case {\it vs.} $M_A$.
The red points are obtained by averaging 
the individual data values in bins containing at least
100 points.  
In the sub-Alfv\'enic range, wave action is seen to increase 
towards $M_A = 1$. 
For $M_A \gtrsim 2$ the magnetic wave action
appears to decrease until about $M_A = 3$, possibly with near constant 
value thereafter. 
Similar behavior is observed for the 
velocity wave action in the 
right panel. Once again this demonstrates 
departures from WKB theory, 
at a level that should be viewed as significant.

\begin{figure}[h]
\includegraphics[width=\columnwidth]{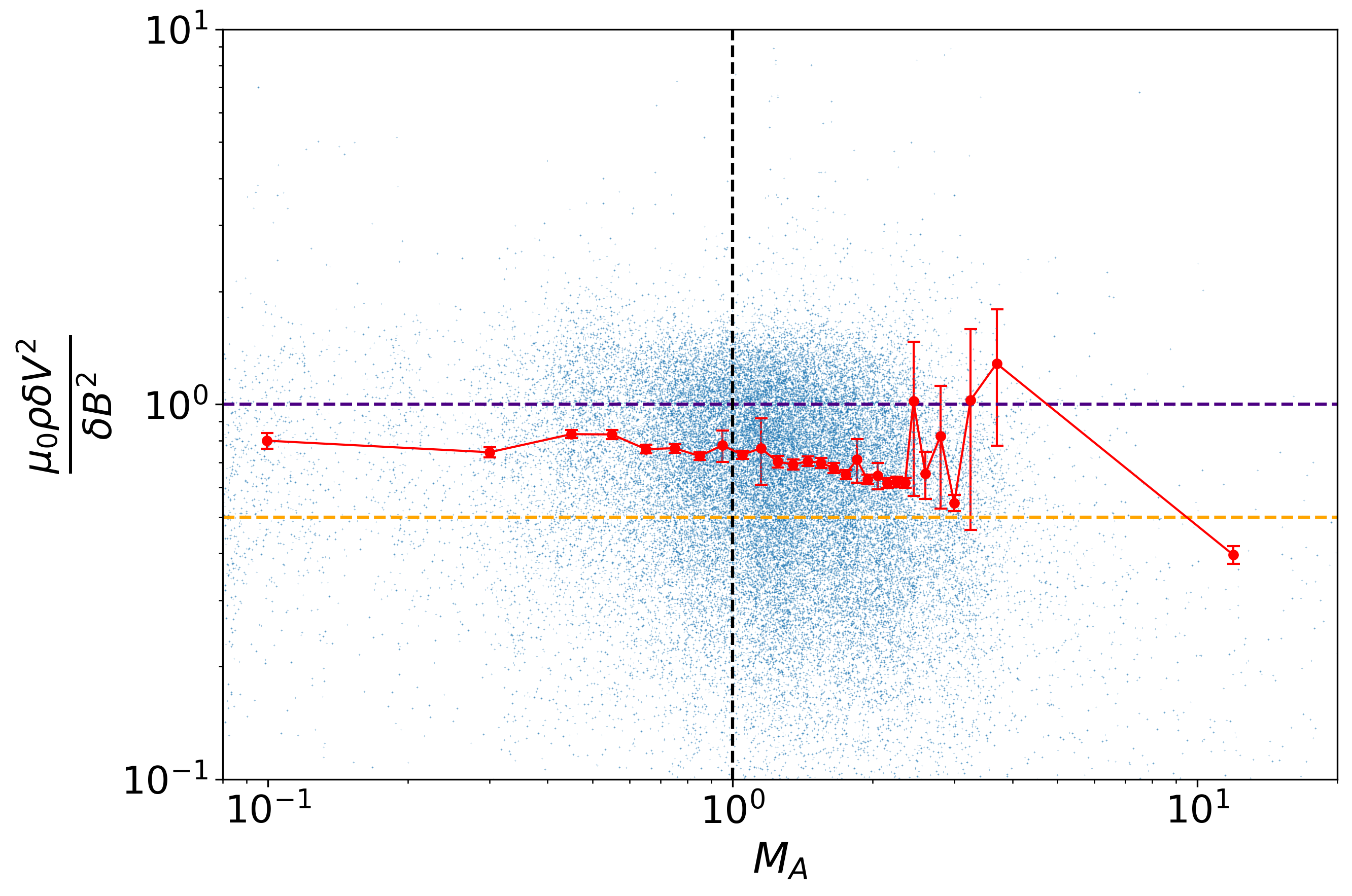}
\caption{ Alfv\'en ratio $r_A = \mu_0\rho\delta V^2 / \delta B^2$ of solar wind fluctuation energies vs.\  Alfv\'en Mach number $M_A$. 
WKB theory for 
non-interacting fossil Alfv\'en waves predicts 
equipartition, i.e., 
$r_A=1$. 
This PSP observation near the Sun is 
consistent 
with observations at greater distances that
typically 
indicate $r_A < 1$.
This 
can be understood in terms of turbulence effects
\citep[see, e.g.,][]{MatthaeusGoldstein82meas,
MatthaeusLamkin86,
ThepthongEA24}. Orange dashed line marks 0.5 on vertical axis.
}
   \label{fig:alven_ratio}
\end{figure}

Finally, in Figure \ref{fig:alven_ratio} we examine the 
WKB assumption of equipartition
by computing the behavior of the Alfv\'en ratio
$r_A = \mu_0 \rho \delta V^2/\delta B^2$.
This quantity is well understood to attain 
values less than unity (usually near 1/2 at $r\sim1$; orange dashed line) 
in the inertial range 
of turbulence as seen in solar wind observations and in 
MHD simulations \citep{MatthaeusLamkin86}. 
But in WKB theory it is expected to have a value of 
unity (blue dashed line). 
Here we test this prediction of wave theory.
From Figure \ref{fig:alven_ratio} we see that the average Alfv\'en ratio within
bins in $M_A$ is inconsistent with equipartition, with $r_A<1$ at all values of 
$M_A$ (except for a few bins with small sample sizes.) This is consistent with typical solar wind observations at larger heliocentric 
distances, near Earth and beyond \citep{MatthaeusGoldstein82meas,BrunoCarboneLRSP13}.
Observation of systematic
averages that 
deviate from 
equipartition represents another departure from 
WKB theory.


\section{Discussion and Summary} \label{sec:discussion}

We have presented here 
what might be considered the first direct observational evidence that WKB wave propagation does not predict the 
radial evolution of 
MHD-scale fluctuations in the sub-Alfv\'enic coronal plasma.
Note that our interpretation focuses on the increase of fluctuation energy far above the trend expected from WKB theory, as the Alfv\'en Mach number increases from $M_A\sim0.5$ to $M_A\sim1$, and does not rely on the more sparse data at $M_A<0.5$. 
We say this is ``direct'' evidence 
because there is a well established class of models that require, at a fundamental 
level, non-WKB effects
to produce 
coronal heating and solar wind acceleration. 
These include {\it reflection driven} and 
{\it wave-turbulence} models
\citep{MattEA99-ch,CranmerEA07,BreechEA08, VerdiniEA10,vanderHolstEA14,UsmanovEA18} that provide reasonable agreement with a variety of observed solar wind properties, even taking into account their subtle differences. 
Quite generally these   
models involve non-WKB transport \citep{ZhouMatt89-nonwkb}, including {\it leading order} production  
of ``inward'' fluctuations
(meaning the minority cross-helicity or Els\"asser species). 
The same models
typically invoke  a von Karman-style dissipation
function that provides plasma heating due to 
nonlinear couplings 
and a turbulent cascade. 
The successes of these models in explaining 
observations throughout the heliosphere and in particular PSP observations
\citep{AdhikariEA20-PSP,ChhiberEA21-heliorad}
{\it indirectly} support 
the underlying assumptions of 
non-WKB transport. 
A summary of such observational departures of super-Alfv\'enic wind from WKB expectations is given by 
\citep{MatthaeusVelli11}.

The present direct observational evidence that WKB theory is not valid in the sub-Alfv\'enic corona, 
along with the 
supporting indications
based on successes of non-WKB transport 
models, lead to the 
conclusion that 
WKB, as a fundamental theoretical
building block, is not defensible, 
either in the corona or in the super-Alfv\'enic solar wind.  
It is at best a crude approximation and
one that misses numerous important observed phenomena.
Furthermore, this shows that non-interacting 
fossil fluctuations that propagate outward from the Sun should rapidly lose energy with increasing $M_A$ and therefore represent at most a small portion of the total fluctuation energy even at the Alfv\'en critical region, which typically occurs at $r= 15$ to $20 R_\sun$.

There do remain nevertheless
numerous 
references in the literature to arguments based on WKB that are purported to explain coronal phenomena.
\cite{PerezChandran13}
use WKB to demonstrate the validity of their coronal model.
\cite{SquireEA20}
attribute the radial behavior of coronal fluctuations to WKB physics. 
In major recent reviews of PSP observations
\citep[e.g.,][]{RaouafiEA23-parker} 
some authors describe alternative models for explaining 
the occurrence of ``switchbacks'' in which the radial magnetic field $B_R$ varies strongly or temporarily reverses in sign.
A number of these models, but not all, depend on WKB-based reasoning. 
It would be an overreach to 
presume that all such applications of WKB
can be discarded as a conceptual elements in constructing physical models. 
The  theory still can occupy a role as an interesting limiting idealized case. It also can happen that WKB-like
behavior emerges somewhat fortuitously, in a more complex scenario, as suggested by 
\citet{VermaRoberts93} in a dissipationless model, 
and as shown to emerge in a 
balance between dissipation and forcing by \citet{ZankEA96}.

It is useful at this point to revisit the 
importance of the paper by 
\citet{HeinemannOlbert80}. 
This seminal work quantified the presence of 
non-WKB coupling between upward- and downward-traveling one-dimensional small amplitude waves, a type of 
``reflection'' caused by spatial variations of Alfv\'en speed (or density). This is an antecedent
of the more complete ``mixing'' term described 
by \citet{ZhouMatt89-nonwkb} and later suggestively called the MECS terms for ``Mixing, Expansion, Compression and Shear'' by \citet{ZankEA96}.
In \citet{HeinemannOlbert80},
reflection occurs when the wave frequency 
become low enough that the WKB scale expansion
begins to fail. This is enough to introduce 
some non-WKB effects, but having a long wavelength is clearly not the only way to enter the non-WKB regime.
The distribution of wavevectors in three dimensions 
and nonlinearity 
are both also important, or even 
dominant factors. 
In the nonlinear regime, fluctuations no longer obey dispersion relations derived in linear wave theory. 
When wave packets with inward- and outward-type polarizations (Els\"asser packets in $z^+$ and $z^-$) have 
high  enough frequency difference, there is no coupling and WKB is enforced. This property
is shown in detail in \cite{MattEA94-wkb}.
In contrast, for strong
turbulence, almost all Fourier amplitudes (in wave vector space) 
have significant power at low, nearly zero frequency \citep{DmitrukMatthaeus09}. The frequency {\it difference} of these colliding fluctuations is so low that WKB cannot be enforced. Reflection, and other MECS effects can then occur. Heinemann and Olbert 
describe one route to such low frequency differences,
but strong turbulence, as well as quasi-two dimensional (2D) turbulence (with very long parallel wavelength)
are equally effective, and perhaps more realistic elements of inertial range solar wind and coronal fluctuations. 

Another interesting element in \citet{HeinemannOlbert80} is the derivation of a conserved {\it total} 
wave action
that applies to their small amplitude one dimensional case when when WKB ordering is not enforced
and oppositely traveling wave packets are coupled
by reflection, but not by local turbulence. 
It may be possible to derive
a more general conservation law of this type when 
considering the 
full structure  of the 
mixing terms in the linear non-WKB 
transport equations \citep[see][]{MattEA94-wkb,OughtonMatt95}.
Such a generalized conservation law
may exhibit an inflection point at $M_A = 1$, as seen
in the Heinemann and Olbert case, due to the change in the net group velocity (including the solar wind speed) of inward-propagating waves, from propagation toward $-\hat r$ for $M_A<1$ to $+\hat r$ for $M_A>1$.
This might help explain the 
inflection at $M_A=1$ as seen from the observations
in our Figure \ref{fig:db-MA-02}.

It should be understood that the mixing terms 
are of crucial importance in causing violations of 
the WKB expansion. But the same terms are also responsible for triggering turbulent cascades in MHD-like plasma flows including those in the corona and 
solar wind. The turbulence, once present, also 
produces the broad range of frequencies at each wavelength 
that prohibits the occurrence of WKB ordering. 
But another significant effect of the 
mixing/MECS terms in non-WKB transport is 
the {\it production} of fluctuation energy. This can occur by 
the conservative exchange of energy 
from large scale shears (or magnetic shears)   
into small-scale fluctuations. 

This, in outline, is the way that nonlinear Kelvin-Helmholtz rollups and
mixing layers in hydrodynamics lead to enhanced turbulence. An analogous pathway for tapping
coronal shear flows 
\citep{DeForestEA18} to form MHD scale rollups and ``switchbacks'' was proposed by \citet{Ruffolo2020}
to occur near and outside the $M_A\sim1$ transition region. 
Indeed the latter work predicted that sub-Alfv\'enic solar wind (which was not yet observed at that time) should have much weaker fluctuation energy, a prediction that we validate here.
It is possible 
that the apparent buildup of
fluctuation energy that we document here, 
as $M_A=1$ is approached from below, 
may represent the onset of this energy 
exchange from coronal shears. 
Observations also indicate that the actual rollups/switchbacks
occur in a more fully developed state
at $M_A>1$ \citep{BandyopadhyayEA22sub, PecoraEA22sb,Jagarlamudi2023ApJ}. Further study will
be needed to more fully understand the physics 
of this fascinating region of the corona and solar wind near the Alfv\'en critical zone \citep{ChhiberEA22,CranmerEA23-SolarPhys}.



\medskip
This research is 
partially supported in Thailand by the National Science and Technology Development Agency (NSTDA) and National Research Council (NRCT): High-Potential Research Team Grant Program (N42A650868), 
and from the NSRF via the Program Management Unit for Human Resources \& Institutional Development, Research and Innovation (B39G670013). It was also supported by 
the NASA 
Heliospheric Supporting Research program (grant 80NSSC18K1648), 
the NASA Parker Solar Probe Guest Investigator program (80NSSC21K1765), the NASA LWS Science program (grant 80NSSC22K1020), 
and 
the Parker Solar Probe mission under the ISOIS project (contract NNN06AA01C) and a subcontract to the University of Delaware from Princeton University. 



\begin{thebibliography}{}
\expandafter\ifx\csname natexlab\endcsname\relax\def\natexlab#1{#1}\fi
\providecommand{\url}[1]{\href{#1}{#1}}
\providecommand{\dodoi}[1]{doi:~\href{http://doi.org/#1}{\nolinkurl{#1}}}
\providecommand{\doeprint}[1]{\href{http://ascl.net/#1}{\nolinkurl{http://ascl.net/#1}}}
\providecommand{\doarXiv}[1]{\href{https://arxiv.org/abs/#1}{\nolinkurl{https://arxiv.org/abs/#1}}}

\bibitem[{{Adhikari} {et~al.}(2020){Adhikari}, {Zank}, \& {Zhao}}]{AdhikariEA20-PSP}
{Adhikari}, L., {Zank}, G.~P., \& {Zhao}, L.~L. 2020, \apj, 901, 102, \dodoi{10.3847/1538-4357/abb132}

\bibitem[{{Badman} {et~al.}(2023){Badman}, {Stevens}, {Paulson}, {Rivera}, {Niembro Hernandez}, {Verniero}, {Livi}, {Larson}, {Kasper}, {Dakeyo}, \& {Riley}}]{BadmanEA2023AGU}
{Badman}, S.~T., {Stevens}, M.~L., {Paulson}, K.~W., {et~al.} 2023, in AGU Fall Meeting Abstracts, Vol. 2023, SH33B--01

\bibitem[{Bale {et~al.}(2016)Bale, Goetz, Harvey, Turin, Bonnell, de~Wit, Ergun, MacDowall, Pulupa, Andre, Bolton, Bougeret, Bowen, Burgess, Cattell, Chandran, Chaston, Chen, Choi, Connerney, Cranmer, Diaz-Aguado, Donakowski, Drake, Farrell, Fergeau, Fermin, Fischer, Fox, Glaser, Goldstein, Gordon, Hanson, Harris, Hayes, Hinze, Hollweg, Horbury, Howard, Hoxie, Jannet, Karlsson, Kasper, Kellogg, Kien, Klimchuk, Krasnoselskikh, Krucker, Lynch, Maksimovic, Malaspina, Marker, Martin, Martinez-Oliveros, McCauley, McComas, McDonald, Meyer-Vernet, Moncuquet, Monson, Mozer, Murphy, Odom, Oliverson, Olson, Parker, Pankow, Phan, Quataert, Quinn, Ruplin, Salem, Seitz, Sheppard, Siy, Stevens, Summers, Szabo, Timofeeva, Vaivads, Velli, Yehle, Werthimer, \& Wygant}]{Bale_2016}
Bale, S.~D., Goetz, K., Harvey, P.~R., {et~al.} 2016, Space science reviews, 204, 49-82, \dodoi{10.1007/s11214-016-0244-5}

\bibitem[{Bandyopadhyay {et~al.}(2022)Bandyopadhyay, Matthaeus, McComas, Chhiber, Usmanov, Huang, Livi, Larson, Kasper, Case, {et~al.}}]{BandyopadhyayEA22sub}
Bandyopadhyay, R., Matthaeus, W., McComas, D., {et~al.} 2022, The Astrophysical journal letters, 926, L1

\bibitem[{{Barnes}(1975)}]{Barnes75}
{Barnes}, A. 1975, Advances in Electronics and Electron Physics, 36, 1, \dodoi{10.1016/S0065-2539(08)61117-8}

\bibitem[{Barnes(1979)}]{Barnes79a}
Barnes, A. 1979, in Solar System Plasma Physics, vol. {I}, ed. E.~N. Parker, C.~F. Kennel, \& L.~J. Lanzerotti (Amsterdam: North-Holland), 251

\bibitem[{Belcher \& Davis(1971)}]{Belcher_Davis_1971}
Belcher, J.~W., \& Davis, Leverett, J. 1971, Journal of geophysical research, 76, 3534-3563, \dodoi{10.1029/ja076i016p03534}

\bibitem[{Belcher \& Davis~Jr.(1971)}]{BelcherDavis71}
Belcher, J.~W., \& Davis~Jr., L. 1971, J.\ Geophys.\ Res., 76, 3534

\bibitem[{Breech {et~al.}(2008)Breech, Matthaeus, Minnie, Bieber, Oughton, Smith, \& Isenberg}]{BreechEA08}
Breech, B., Matthaeus, W.~H., Minnie, J., {et~al.} 2008, J.\ Geophys.\ Res., 113, \dodoi{10.1029/2007JA012711}

\bibitem[{{Bruno} \& {Carbone}(2013)}]{BrunoCarboneLRSP13}
{Bruno}, R., \& {Carbone}, V. 2013, Living Reviews in Solar Physics, 10, 2, \dodoi{10.12942/lrsp-2013-2}

\bibitem[Chhiber(2022)]{Chhiber22} Chhiber, R.\ 2022, \apj, 939, 33. doi:10.3847/1538-4357/ac9386

\bibitem[{{Chhiber} {et~al.}(2022){Chhiber}, {Matthaeus}, {Usmanov}, {Bandyopadhyay}, \& {Goldstein}}]{ChhiberEA22}
{Chhiber}, R., {Matthaeus}, W.~H., {Usmanov}, A.~V., {Bandyopadhyay}, R., \& {Goldstein}, M.~L. 2022, \mnras, 513, 159, \dodoi{10.1093/mnras/stac779}

\bibitem[{{Chhiber} {et~al.}(2021){Chhiber}, {Usmanov}, {Matthaeus}, \& {Goldstein}}]{ChhiberEA21-heliorad}
{Chhiber}, R., {Usmanov}, A.~V., {Matthaeus}, W.~H., \& {Goldstein}, M.~L. 2021, The Astrophysical Journal, 923, 89

\bibitem[{{Chhiber} {et~al.}(2024){Chhiber}, {Pecora}, {Usmanov}, {Matthaeus}, {Goldstein}, {Roy}, {Wang}, {Thepthong}, \& {Ruffolo}}]{ChhiberEA24}
{Chhiber}, R., {Pecora}, F., {Usmanov}, A.~V., {et~al.} 2024, \mnras, 533, L70, \dodoi{10.1093/mnrasl/slae051}

\bibitem[{Coleman~Jr(1967)}]{Coleman67wave}
Coleman~Jr, P.~J. 1967, Planetary and Space Science, 15, 953

\bibitem[{Cranmer {et~al.}(2007)Cranmer, {van}~{Ballegooijen}, \& Edgar}]{CranmerEA07}
Cranmer, S.~R., {van}~{Ballegooijen}, A.~A., \& Edgar, R.~J. 2007, Astrophys.\ J.\ Suppl. Ser, 171, 520, \dodoi{10.1086/518001}

\bibitem[{Cranmer {et~al.}(2023)Cranmer, Chhiber, Gilly, Cairns, Colaninno, McComas, Raouafi, Usmanov, Gibson, \& DeForest}]{CranmerEA23-SolarPhys}
Cranmer, S.~R., Chhiber, R., Gilly, C.~R., {et~al.} 2023, Solar Physics, 298, 126

\bibitem[{{DeForest} {et~al.}(2018){DeForest}, {Howard}, {Velli}, {Viall}, \& {Vourlidas}}]{DeForestEA18}
{DeForest}, C.~E., {Howard}, R.~A., {Velli}, M., {Viall}, N., \& {Vourlidas}, A. 2018, \apj, 862, 18, \dodoi{10.3847/1538-4357/aac8e3}

\bibitem[{{Dmitruk} \& {Matthaeus}(2009)}]{DmitrukMatthaeus09}
{Dmitruk}, P., \& {Matthaeus}, W.~H. 2009, Physics of Plasmas, 16, 062304, \dodoi{10.1063/1.3148335}

\bibitem[Fargette et al.(2022)]{Fargette22} Fargette, N., Lavraud, B., Rouillard, A.~P., et al.\ 2022, \aap, 663, A109. doi:10.1051/0004-6361/202243537

\bibitem[{{Fisk} \& {Kasper}(2020)}]{FiskKasper20}
{Fisk}, L.~A., \& {Kasper}, J.~C. 2020, \apjl, 894, L4, \dodoi{10.3847/2041-8213/ab8acd}

\bibitem[{Heinemann \& Olbert(1980)}]{HeinemannOlbert80}
Heinemann, M., \& Olbert, S. 1980, J.\ Geophys.\ Res., 85, 1311, \dodoi{10.1029/JA085iA03p01311}

\bibitem[{Hollweg(1974)}]{Hollweg74}
Hollweg, J.~V. 1974, J.\ Geophys.\ Res., 79, 1539

\bibitem[{Hollweg(1990)}]{Hollweg90}
---. 1990, J.\ Geophys.\ Res., 95, 14\,873

\bibitem[{Jacques(1977)}]{Jacques77}
Jacques, S.~A. 1977, Astrophys.\ J., 215, 942

\bibitem[{{Jacques}(1978)}]{Jacques78}
{Jacques}, S.~A. 1978, \apj, 226, 632, \dodoi{10.1086/156647}

\bibitem[{{Jagarlamudi} {et~al.}(2023){Jagarlamudi}, {Raouafi}, {Bourouaine}, {Mostafavi}, {Larosa}, \& {Perez}}]{Jagarlamudi2023ApJ}
{Jagarlamudi}, V.~K., {Raouafi}, N.~E., {Bourouaine}, S., {et~al.} 2023, \apjl, 950, L7, \dodoi{10.3847/2041-8213/acd778}

\bibitem[{Kasper {et~al.}(2016)Kasper, Abiad, Austin, Balat-Pichelin, Bale, Belcher, Berg, Bergner, Berthomier, Bookbinder, Brodu, Caldwell, Case, Chandran, Cheimets, Cirtain, Cranmer, Curtis, Daigneau, Dalton, Dasgupta, DeTomaso, Diaz-Aguado, Djordjevic, Donaskowski, Effinger, Florinski, Fox, Freeman, Gallagher, Gary, Gauron, Gates, Goldstein, Golub, Gordon, Gurnee, Guth, Halekas, Hatch, Heerikuisen, Ho, Hu, Johnson, Jordan, Korreck, Larson, Lazarus, Li, Livi, Ludlam, Maksimovic, McFadden, Marchant, Maruca, McComas, Messina, Mercer, Park, Peddie, Pogorelov, Reinhart, Richardson, Robinson, Rosen, Skoug, Slagle, Steinberg, Stevens, Szabo, Taylor, Tiu, Turin, Velli, Webb, Whittlesey, Wright, Wu, \& Zank}]{Kasper_2016}
Kasper, J.~C., Abiad, R., Austin, G., {et~al.} 2016, Space science reviews, 204, 131-186, \dodoi{10.1007/s11214-015-0206-3}

\bibitem[{{Livi} {et~al.}(2022){Livi}, {Larson}, {Kasper}, {Abiad}, {Case}, {Klein}, {Curtis}, {Dalton}, {Stevens}, {Korreck}, {Ho}, {Robinson}, {Tiu}, {Whittlesey}, {Verniero}, {Halekas}, {McFadden}, {Marckwordt}, {Slagle}, {Abatcha}, {Rahmati}, \& {McManus}}]{Livi2022}
{Livi}, R., {Larson}, D.~E., {Kasper}, J.~C., {et~al.} 2022, \apj, 938, 138, \dodoi{10.3847/1538-4357/ac93f5}

\bibitem[{{Matthaeus} \& {Goldstein}(1982)}]{MatthaeusGoldstein82meas}
{Matthaeus}, W.~H., \& {Goldstein}, M.~L. 1982, \jgr, 87, 6011, \dodoi{10.1029/JA087iA08p06011}

\bibitem[{Matthaeus \& Lamkin(1986)}]{MatthaeusLamkin86}
Matthaeus, W.~H., \& Lamkin, S.~L. 1986, The Physics of fluids, 29, 2513, \dodoi{10.1063/1.866004}

\bibitem[{Matthaeus {et~al.}(1994{\natexlab{a}})Matthaeus, Oughton, Pontius, \& Zhou}]{MattEA94-ec}
Matthaeus, W.~H., Oughton, S., Pontius, D., \& Zhou, Y. 1994{\natexlab{a}}, J.\ Geophys.\ Res., 99, 19\,267

\bibitem[{{Matthaeus} \& {Velli}(2011)}]{MatthaeusVelli11}
{Matthaeus}, W.~H., \& {Velli}, M. 2011, Space Sci. Rev., 160, 145, \dodoi{10.1007/s11214-011-9793-9}

\bibitem[{Matthaeus {et~al.}(1999)Matthaeus, Zank, Oughton, Mullan, \& Dmitruk}]{MattEA99-ch}
Matthaeus, W.~H., Zank, G.~P., Oughton, S., Mullan, D.~J., \& Dmitruk, P. 1999, Astrophys.\ J., 523, L93

\bibitem[{Matthaeus {et~al.}(1994{\natexlab{b}})Matthaeus, Zhou, Zank, \& Oughton}]{MattEA94-wkb}
Matthaeus, W.~H., Zhou, Y., Zank, G.~P., \& Oughton, S. 1994{\natexlab{b}}, J.\ Geophys.\ Res., 99, 23\,421

\bibitem[{Moncuquet {et~al.}(2020)Moncuquet, Meyer-Vernet, Issautier, Pulupa, Bonnell, Bale, de~Wit, Goetz, Griton, Harvey, MacDowall, Maksimovic, \& Malaspina}]{Moncuquet_2020}
Moncuquet, M., Meyer-Vernet, N., Issautier, K., {et~al.} 2020, The Astrophysical journal. Supplement series, 246, 44, \dodoi{10.3847/1538-4365/ab5a84}

\bibitem[{Oughton \& Matthaeus(1995)}]{OughtonMatt95}
Oughton, S., \& Matthaeus, W.~H. 1995, J.\ Geophys.\ Res., 100, 14\,783

\bibitem[{Oughton {et~al.}(2015)Oughton, Matthaeus, Wan, \& Osman}]{OughtonEA15}
Oughton, S., Matthaeus, W.~H., Wan, M., \& Osman, K.~T. 2015, Philosophical Transactions of the Royal Society A, 373, 20140152, \dodoi{10.1098/rsta.2014.0152}

\bibitem[{{Parker}(1965)}]{Parker65-ssr}
{Parker}, E.~N. 1965, \ssr, 4, 666, \dodoi{10.1007/BF00216273}

\bibitem[{Pecora {et~al.}(2022)Pecora, Matthaeus, Primavera, Greco, Chhiber, Bandyopadhyay, \& Servidio}]{PecoraEA22sb}
Pecora, F., Matthaeus, W.~H., Primavera, L., {et~al.} 2022, The Astrophysical Journal Letters, 929, L10

\bibitem[{Perez \& Chandran(2013)}]{PerezChandran13}
Perez, J.~C., \& Chandran, B. D.~G. 2013, The Astrophysical Journal, 776, 124, \dodoi{10.1088/0004-637X/776/2/124}

\bibitem[{Raouafi {et~al.}(2023)Raouafi, Matteini, Squire, Badman, Velli, Klein, Chen, Matthaeus, Szabo, Linton, {et~al.}}]{RaouafiEA23-parker}
Raouafi, N.~E., Matteini, L., Squire, J., {et~al.} 2023, Space Science Reviews, 219, 8

\bibitem[{Romeo {et~al.}(2023)Romeo, Braga, Badman, Larson, Stevens, Huang, Phan, Rahmati, Livi, Alnussirat, {et~al.}}]{romeo2023}
Romeo, O., Braga, C., Badman, S., {et~al.} 2023, The Astrophysical Journal, 954, 168, \dodoi{https://doi.org/10.3847/1538-4357/ace62e}

\bibitem[{{Ruffolo} {et~al.}(2020){Ruffolo}, {Matthaeus}, {Chhiber}, {Usmanov}, {Yang}, {Bandyopadhyay}, {Parashar}, {Goldstein}, {DeForest}, {Wan}, {Chasapis}, {Maruca}, {Velli}, \& {Kasper}}]{Ruffolo2020}
{Ruffolo}, D., {Matthaeus}, W.~H., {Chhiber}, R., {et~al.} 2020, \apj, 902, 94, \dodoi{10.3847/1538-4357/abb594}

\bibitem[{{Squire} {et~al.}(2020){Squire}, {Chandran}, \& {Meyrand}}]{SquireEA20}
{Squire}, J., {Chandran}, B.~D.~G., \& {Meyrand}, R. 2020, \apjl, 891, L2, \dodoi{10.3847/2041-8213/ab74e1}

\bibitem[{{Thepthong} {et~al.}(2024){Thepthong}, {Pongkitiwanichakul}, {Ruffolo}, {Kieokaew}, {Bandyopadhyay}, {Matthaeus}, \& {Parashar}}]{ThepthongEA24}
{Thepthong}, P., {Pongkitiwanichakul}, P., {Ruffolo}, D., {et~al.} 2024, \apj, 962, 37, \dodoi{10.3847/1538-4357/ad1592}

\bibitem[{{Usmanov} {et~al.}(2018){Usmanov}, {Matthaeus}, {Goldstein}, \& {Chhiber}}]{UsmanovEA18}
{Usmanov}, A.~V., {Matthaeus}, W.~H., {Goldstein}, M.~L., \& {Chhiber}, R. 2018, Astrophys. J., 865, 25, \dodoi{10.3847/1538-4357/aad687}

\bibitem[{{van der}~{Holst} {et~al.}(2014){van der}~{Holst}, Sokolov, Meng, Jin, Manchester, T{\'o}th, \& Gombosi}]{vanderHolstEA14}
{van der}~{Holst}, B., Sokolov, I.~V., Meng, X., {et~al.} 2014, Astrophys.\ J., 782, 81, \dodoi{10.1088/0004-637X/782/2/81}

\bibitem[{{Verdini} {et~al.}(2010){Verdini}, {Velli}, {Matthaeus}, {Oughton}, \& {Dmitruk}}]{VerdiniEA10}
{Verdini}, A., {Velli}, M., {Matthaeus}, W.~H., {Oughton}, S., \& {Dmitruk}, P. 2010, Astrophys. J. Lett., 708, L116, \dodoi{10.1088/2041-8205/708/2/L116}

\bibitem[{Verma \& Roberts(1993)}]{VermaRoberts93}
Verma, M.~K., \& Roberts, D.~A. 1993, J. Geophys. Res., 98, 5625

\bibitem[{Wan {et~al.}(2012)Wan, Oughton, Servidio, \& Matthaeus}]{WanEA12-jfm}
Wan, M., Oughton, S., Servidio, S., \& Matthaeus, W.~H. 2012, J. Fluid Mech., 697, 296, \dodoi{10.1017/jfm.2012.61}

\bibitem[{Weinberg(1962)}]{Weinberg62}
Weinberg, S. 1962, Phys.\ Rev., 126, 1899

\bibitem[{{Zank} {et~al.}(1996){Zank}, {Matthaeus}, \& {Smith}}]{ZankEA96}
{Zank}, G.~P., {Matthaeus}, W.~H., \& {Smith}, C.~W. 1996, \jgr, 101, 17093, \dodoi{10.1029/96JA01275}

\bibitem[{Zhou \& Matthaeus(1989)}]{ZhouMatt89-nonwkb}
Zhou, Y., \& Matthaeus, W.~H. 1989, Geophysical Research Letters, 16, 755

\end{thebibliography}
\bibliographystyle{aasjournal}
 \newcommand{\BIBand} {and} 



\end{document}